\documentstyle[aps,prl,epsf,epsfig]{revtex}
\begin{document}
\draft
\twocolumn[
\hsize\textwidth\columnwidth\hsize\csname @twocolumnfalse\endcsname

\title{
Ordering kinetics of stripe patterns
}
\author{
Denis Boyer and Jorge Vi\~nals
}
\address{ 
School of Computational Science and Information Technology,\\
Florida State University, Tallahassee, Florida 32306-4120.
}
\date{\today}
\maketitle
\begin{abstract}

We study domain coarsening of two dimensional stripe patterns
by numerically solving the Swift-Hohenberg model of Rayleigh-B\'enard
convection. Near the bifurcation threshold, the evolution of disordered 
configurations is dominated by grain boundary motion through a background of 
largely immobile curved stripes.
A numerical study of the distribution of local stripe curvatures, of the 
structure factor of the order parameter, and a finite size scaling analysis 
of the grain boundary perimeter, suggest that the linear scale of the structure 
grows as a power law of time $t^{1/z}$, with $z=3$. 
We 
interpret theoretically
the exponent $z=3$ from the law of grain boundary motion.

\end{abstract}

\narrowtext
]

Equilibrium layered phases (characterized by a uniform 
wavevector $\vec{k}_{0} \neq 0$) are often found in systems with 
competing short and long ranged interactions \cite{re:seul95}. 
Related structures, commonly referred to as stripe patterns, 
also appear in systems driven outside of thermodynamic equilibrium 
(e.g., Rayleigh-B\'enard convection or 
parametric surface waves near onset \cite{re:cross93}).
After changing rapidly a control parameter across a transition or bifurcation 
point, a uniform state become unstable and configurations with locally ordered
stripes appear. Given the underlying translational and rotational
invariances of the system, spontaneous evolution leads to
a macroscopic sample comprising a large number of grains or domains, each
relatively uniform, but oriented along an arbitrary direction, 
as well as to a large
density of defects such as grain boundaries, disclinations and dislocations.
Understanding how this structure orders 
with time, and how the motion of interacting defects contributes to the 
coarsening rate is the main focus of this paper.

Numerical studies of model equations in {\em two} spatial dimensions 
\cite{re:elder92,re:elder92b,re:cross95a,re:hou97,re:christensen98},
as well as recent experiments involving thin films of block copolymers 
\cite{re:harrison00b} support the idea that the time evolution of 
layered phases after a quench is statistically self-similar
(the statistical self-similarity hypothesis asserts that after a possible
transient, consecutive configurations 
of the coarsening structure are geometrically similar in a statistical sense). 
As a consequence, any linear scale of the structure (e.g., the average 
size of a domain or grain) is expected to grow as a power law of time 
$R(t) \sim t^{1/z}$.

Coarsening of layered phases is not yet well understood.
On symmetry grounds, layered phases can be classified as
smectics \cite{re:toner81,re:elder92b}.
Hence, by analogy with coarsening studies of nematics \cite{re:zapotocky95} 
and O(N) vector models with a nonconserved order parameter \cite{re:mazenko95}, 
one would argue that self-similar coarsening is to be expected with $z=2$.  
Although, the possibility of a long time cross over to $z=2$ has in fact been 
considered \cite{re:elder92b}, numerical evidence has consistently pointed 
at values of $z$ in the range $z=4-5$
\cite{re:elder92b,re:cross95a,re:hou97,re:christensen98}. 
More importantly, the self-similarity hypothesis itself has been questioned as 
different linear scales yield different values of $z$
\cite{re:hou97,re:christensen98}.
Furthermore, and in contrast with related research on well understood
systems that order at $k_{0} = 0$ \cite{re:gunton83,re:bray94}, 
the value of $z$ appears to be modified by the presence of thermal noise.

We present here a numerical investigation of domain coarsening for the
Swift-Hohenberg model of Rayleigh-B\'enard convection \cite{re:swift77}.
Briefly, we focus on the region close to onset ($\epsilon \rightarrow 0$,
where $\epsilon$ is the reduced control parameter)) and 
find that coarsening 
proceeds in a self-similar manner. We analyze several different 
characteristic length scales, and find that they are asymptotically 
proportional to each other. We also find that $z \simeq 3$, independent of 
thermal noise. We interpret this value of $z$ from
the law of grain boundary motion given in \cite{re:boyer01}.
Further from onset ($\epsilon\simeq 0.25$),
we recover the results of previous studies, which we interpret as arising from
non-adiabatic effects that lead to defect pinning. This fact accounts for
both the slower growth seen previously, and its dependence on fluctuations.

The Swift-Hohenberg model of convection in dimensionless units is
\cite{re:swift77},
\begin{equation}
\label{sh}
\frac{\partial \psi}{\partial t}=
\epsilon\psi-\frac{\xi_{0}^{2}}{4k_0^2} (k_0^2+\nabla^2)^2\psi-\psi^3\ ,
\end{equation}
where $\psi$ is an order parameter related to the vertical fluid velocity at 
the mid plane of a Rayleigh-B\'enard convection cell, $\epsilon$ is the reduced
Rayleigh number, $k_{0}$ is the roll wavenumber, and
$\xi_{0}$ is a constant that depends on the boundary conditions at the top and
bottom plates. For our purposes, we note that $\xi_{0} \propto 1/k_{0}$. The
same model has been used to analyze coarsening of lamellar phases in a diblock 
copolymer \cite{re:christensen98}, and is otherwise believed to be a
generic model of the kinetics of stripe formation. There, $\psi$ is the 
concentration 
difference between the two monomers \cite{re:shiwa97,re:hohenberg95}. For
$0<\epsilon \ll 1$, the stationary solution of
Eq. (\ref{sh}) is well approximated by a sinusoidal function of wavenumber 
$k_{0}$. The transient evolution and domain coarsening is investigated 
by numerically integrating Eq. (\ref{sh}) from random initial
conditions. All calculations are performed very close to onset 
($\epsilon=0.04$). Details of the numerical algorithm can be found in ref. 
\cite{re:boyer01}.
\begin{figure}
\epsfig{figure=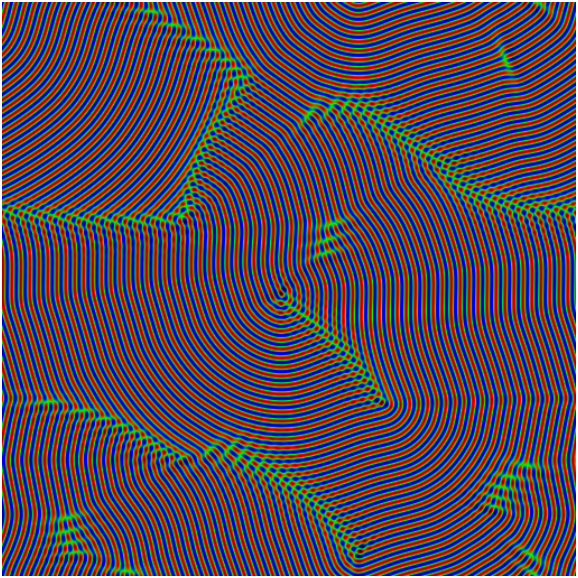,width=3in}
\vspace{0.2cm}
\caption{Order parameter $\psi$ shown in grey scale at time $t = 15000$. 
Eq. (\ref{sh}) is discretized on a square grid of mesh size $\Delta x = 1$ 
with $512^2$ nodes. The wavelength $\lambda_0= 2\pi/k_0 = 8\Delta x$. The 
reduced Rayleigh number is 
$\epsilon = 0.04$. The initial condition has $\langle \psi \rangle = 0$ 
and $\langle \psi^{2} \rangle =0.04$.}
\label{figanim}
\end{figure}
Figure \ref{figanim} shows a typical transient configuration.
The configuration contains a large amount of grain boundaries that
separate domains of different orientation, as well as defects 
(such as +1/2 disclinations and dislocations).

We first present our numerical results for several measures of the linear
scale of the structure, including the distribution of stripe curvatures, the
order parameter structure factor, and the grain boundary perimeter.
Following initial transients, the stripe curvature
$\kappa = |\nabla\cdot\vec{n}|$ is a slowly varying field, where 
$\vec{n}$ denotes the unitary vector normal to the lines of constant $\psi$. 
We compute the probability distribution function of stripe curvatures 
$P(\kappa,t)$ by considering only the subset of points where
stripe orientation can be properly defined, {\it i.e.} the points that are 
not in the immediate vicinity of any grain boundary nor other defects.
Far enough from a defect,
$\psi(\vec{r})=A(\vec{r})\cos(\vec{k}(\vec{r})\cdot\vec{r}+\phi)$, with $A$ a
slowly varying amplitude. By defining $\zeta(\vec{r}) =
\psi^2+(\vec{\nabla}\psi)^2/k_0^2$, one has $\zeta(\vec{r})\simeq A^2$.
Note that for stationary parallel stripes of wavenumber 
$k_0$, $\zeta(\vec{r}) = \zeta_0 = 4\epsilon/3$ \cite{re:cross93}. We now 
define defect free regions as those that satisfy
$r_m < \zeta/\zeta_0 < r_M$, with $r_m=0.95$, $r_M=1.05$ (filter $a$)
or $r_m=0.97$, $r_M=1.10$ (filter $b$). We have numerically verified 
that the values of $\zeta$ corresponding to a set of 
moderately curved stripes along their transverse direction remain completely
within the intervals defined by both filters $a$ and $b$. By contrast,
most values  of $\zeta$ in the vicinity of a grain boundary are lower than 
$0.90$. 
\begin{figure}
\epsfig{figure=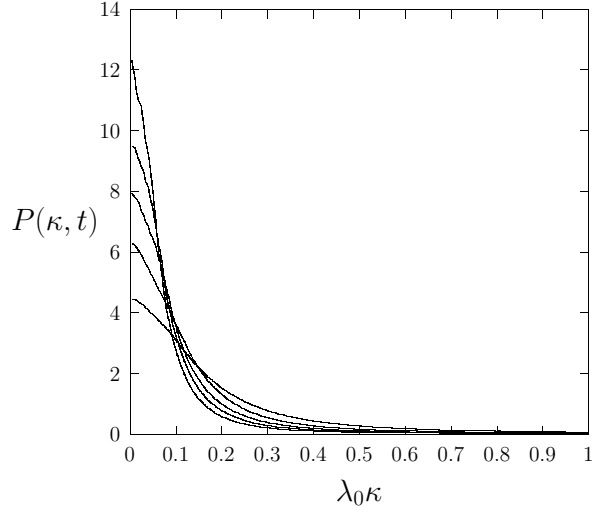,width=3in,bbllx=112,bblly=430,bburx=432,bbury=722}
\caption{Probability distribution function of curvatures $P(\kappa,t)$ averaged
over 35 independent initial conditions for times ranging from $t=960$ to 
$1.6\ 10^4$.}
\label{figpdf}
\end{figure}
\begin{figure}
\epsfig{figure=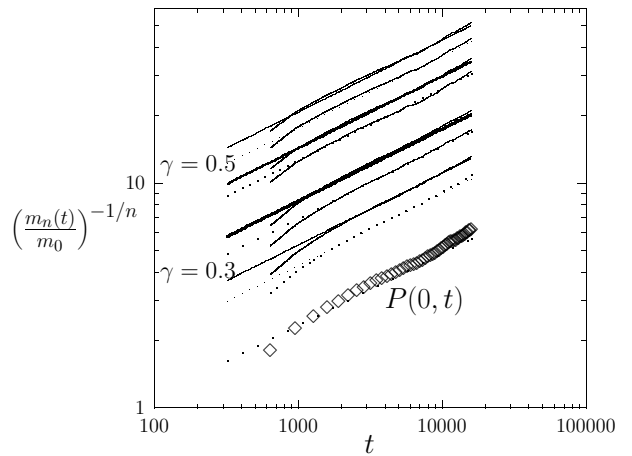,width=3in,bbllx=112,bblly=430,bburx=452,bbury=732}
\caption{Moments $m_{n}(t)$ of $P(\kappa,t)$ obtained 
with filter $a$. The values of $n$ shown are 1/2,1,2,3, from top to bottom. 
The straight lines have a slope of $0.32$.} 
\label{figcurvscal}
\end{figure}
Figure \ref{figpdf} shows $P(\kappa,t)$ at different times, after 
averaging over 35 independent initial conditions on a square grid of size 
$1024 \times 1024$ ($16$ grid nodes per wavelength $\lambda_0$), at 
$\epsilon=0.04$ and using filter $a$. 
To check for self-similarity we study whether $P(\kappa,t)=
t^{1/z} p(\kappa t^{1/z})$. In order to do this, we compute its moments 
$m_n(t)=\int_0^{\kappa_{c}(t)} d\kappa\ \kappa^n P(\kappa,t)$ with
$\kappa_{c}(t)$ defined as
$\int_0^{\kappa_{c}(t)}d\kappa\ P(\kappa,t)=\gamma \kappa_c(t)P(0,t)$,
and $\gamma$ an arbitrary constant, $0<\gamma<1$.
We find that $(m_n/m_0(t))^{-1/n} \propto t^{1/z}$, with the value of $z$
independent of $n$ and $\gamma$, thus lending support to the
self-similarity hypothesis.
Figure \ref{figcurvscal} shows the results for a few values of $n$ and two
values of $\gamma$. In addition, the best fit to the curves yields 
$1/z \simeq 0.32$ with filter $a$, and $1/z \simeq 0.34$ with filter $b$ 
(not shown).

\begin{figure}
\epsfig{figure=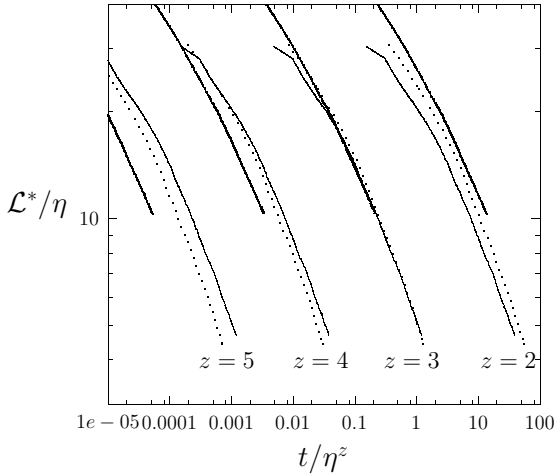,width=3in,bbllx=112,bblly=430,bburx=452,bbury=732}
\caption{Finite size scaling analysis of the total grain boundary
length, with systems of aspect ratio $\eta=32$, $42.66$ and $64$.} 
\label{figfinscal}
\end{figure}
We next present a finite size scaling analysis to independently determine the
value of $z$ \cite{re:vinals88}. 
Let $N_{d}$ be the number of grid points for which $\zeta>r_M$ or $\zeta<r_m$.
The probability of a point belonging to a defect is $p_d=N_d\Delta x^2/L^2$, 
with $L$ the system linear extent. We define a dimensionless defect 
({\it i.e.} grain boundary) perimeter as ${\cal L}^*=\eta^2 p_d$, 
where $\eta=L/\lambda_0$ 
is the aspect ratio. For short times finite size effects are expected to be 
negligible, and $p_d \sim t^{-1/z}$. 
We now introduce a finite size scaling ansatz, valid for any time $t$,
\begin{equation}
\label{finscal}
{\cal L}^{*}(\eta,t) = \eta\ g(t/\eta^z) ,
\end{equation}
with $g(x)\sim x^{-1/z}$ for $x\ll 1$.
At fixed $\epsilon=0.04$, we have
numerically computed ${\cal L}^*(t)$ with the help of 
filter $a$ for different aspect ratios: $\eta=32$, $42.66$ and $64$ 
({\it i.e.} $\eta = 258/8, 256/6$ and $512/8$; $\Delta x=1$ in all cases).
We have averaged the results over $500$, $300$ and $100$ independent initial
conditions respectively. Figure \ref{figfinscal} shows our results for 
the universal curve $g(x)$. The value $z=3$ leads to curves 
${\cal L}^{*}/\eta$ as a function of $t/\eta^z$ do not depend on $\eta$.

We have also analyzed the Fourier transform of the two-point correlation 
function of the order parameter, $S(k,t) = \langle
\tilde{\psi}_{\vec{k}}\tilde{\psi}_{-\vec{k}}\rangle$, averaged over all 
possible orientations of $\vec{k}$ and 50 independent initial conditions. 
Such study is standard \cite{re:elder92b,re:cross95a,re:hou97}. 
If $|k-k_0|\ll k_0$ and $k_0 L(t)\gg 1$, $S(k,t)$ satisfies the 
scaling form $S(k,t)=k_0^{1-d}L(t)\ f[(k^2-k_0^2)L(t)\lambda_0]$,
where $d$ the spatial dimension. Analysis of the moments of $S(k,t)$ 
shows that $L(t)\sim t^{1/z}$ with $1/z\simeq 0.31$. We also find that 
$S(k_0,t) \sim t^{0.32}$. 

Finally, we have verified that the value of the exponent $z$ 
calculated from either the grain boundary perimeter or the structure factor is
not modified by the introduction of random fluctuations into Eq. (\ref{sh}).

The results presented are qualitatively
modified further from onset. For example, at $\epsilon=0.25$
we find that different linear scales of the structure
are no longer proportional to each other, and we obtain
effective exponents that are in agreement with the values of $z$
reported in earlier studies
at $\epsilon\simeq 0.25$ \cite{re:elder92,re:cross95a,re:hou97}.
We find $1/z \simeq 0.21$ from an analysis of the moments of $S(k,t)$,
$1/z \simeq 0.26$ from the grain boundary
perimeter, while the moments of the distribution of curvatures yield
$1/z \simeq 0.32$.

We next discuss a possible growth mechanism leading to an exponent of $z=3$, 
as well as our interpretation for the slower growth that is found when 
$\epsilon$ is not
sufficiently small, and its dependence on random fluctuations. 
Coarsening exponents can be often inferred from the 
law of motion of the class of defects that control coarsening \cite{re:bray98}.
A typical transient configuration (Fig. \ref{figanim})
contains a large amount of grain boundaries, 
as well as other defects such as +1/2 disclinations. Grain boundaries move over
large distances, whereas disclinations remain largely immobile. Disclinations 
produce a background of curved rolls that cannot freely relax due to 
topological constraints (this is in contrast with other coarsening mechanisms 
discussed in \cite{re:elder92,re:christensen98}.)
However, at any given point roll relaxation as well as disclination 
annihilation do occur after the passage of a grain boundary. 
In ref. \cite{re:boyer01} we studied the
motion of a grain boundary separating two semi-infinite domains of mutually
perpendicular rolls, straight on one side, curved on the other. At lowest 
order in
$\epsilon$, those parallel to the boundary are distorted whereas those that are
perpendicular to it remain straight. The energy of the configuration 
decreases by
a net displacement of the grain boundary, the effect of which is to replace 
curved rolls by straight ones of lower energy. Therefore the size of the 
domain with curved
rolls decreases. We showed that if the curvature of the 
rolls ahead of the boundary is $\kappa$, the 
boundary advances at a speed, 
\begin{equation}
\label{eq:scalv}
v_{gb} \sim \epsilon^{-1/2}\kappa^2 .
\end{equation}
It was shown in \cite{re:boyer01}
that Eq. (\ref{eq:scalv}) is in quantitative agreement with a direct numerical 
solution of Eq. (\ref{sh}) with an initial condition that involves a 
90$^{\circ}$ grain boundary. As seen in Fig. \ref{figanim}, disclinations produce
roughly axisymmetric patterns, with a characteristic roll curvature that is
inversely proportional to the distance among disclinations. If this
distance is proportional to the characteristic linear scale $R(t)$, then
dimensional analysis of Eq. (\ref{eq:scalv}) suggests $R(t) \sim t^{1/3}$.

These considerations are modified further from onset.
The law of grain boundary motion, Eq. (\ref{eq:scalv}), is valid only to first 
order in $\epsilon$. At moderate values of $\epsilon$, the
separation of length scales assumed in the derivation of Eq. (\ref{eq:scalv}) 
breaks down (the grain boundary thickness is of order
$\lambda_{0}/\sqrt{\epsilon}$), leading to non-adiabatic effects.
Within the amplitude equation formalism, and following the approach of
ref. \cite{re:bensimon88b}, we have obtained
the leading order nonadiabatic corrections, and find that
the position of the grain boundary and the
phase of the rolls located ahead of it no longer decouple. 
Equation (\ref{eq:scalv}) can be generalized to,
\begin{equation}
\label{motionlaw}
v_{gb}=\frac{\epsilon}{3k_0^2D(\epsilon)}\ \kappa^2\
- \frac{p(\epsilon)}{D(\epsilon)}\cos(2k_0 x_{gb}+\phi)\ ,
\end{equation}
where $x_{bg}$ is the average location of the grain boundary, 
$\phi$ is a constant phase and,
\begin{equation}
D(\epsilon) =\int_{-\infty}^{\infty}
dx\ [(\partial_x A_0)^2+(\partial_x B_0)^2],
\label{D}
\end{equation}
$$
p(\epsilon) ={\max}_{\theta}\left\{
\frac{3}{4}\int_{-\infty}^{\infty} dx\ A_0^3(x)
\partial_xA_0(x)\cos(2k_0x+\theta) \right.
$$
\begin{equation}
\left. + \frac{3}{2}\int_{-\infty}^{\infty} dx\ [2A_0B_0^2\partial_xA_0+
A_0^2B_0 \partial_xB_0]\cos(2k_0x+\theta)\right\}. 
\label{p}
\end{equation}
The functions $A_{0}(x)$ and $B_{0}(x)$ are the amplitudes of the two sets of
rolls separating a planar boundary \cite{re:manneville83b},
the coefficient $D(\epsilon)$ represents a friction term, and $p(\epsilon)$ the
amplitude of a periodic pinning potential. The contribution from non 
adiabatic effects is typically of the order of,
\begin{equation}\label{estimp}
p(\epsilon)\ \sim \ \epsilon^2\ e^{-|\alpha|/\sqrt{\epsilon}}\ ,
\end{equation}
where $|\alpha|$ is a constant of order unity. Hence, $p$ behaves non
analytically as $\epsilon \rightarrow 0$, but increases extremely
fast with $\epsilon$ at low values of $\epsilon$. From Eq. (\ref{motionlaw}) we
see that at any finite $\epsilon>0$, 
there exists a critical curvature $\kappa_g$ below which $v_{gb}=0$.
Remarkably, pinning becomes noticeable even at 
$\epsilon = 0.1$ \cite{re:boyer01}, and grain boundaries were seen to 
advance only by half-integer multiples of the roll wavelength. Therefore 
we expect that grain 
boundaries in a coarsening configuration will become pinned over
time. We believe that this pinning is the reason behind the lower 
effective exponents found in previous studies at $\epsilon = 0.25$ 
\cite{re:elder92,re:hou97,re:christensen98}, as well as for the related result
that random fluctuations added to Eq. (\ref{sh}) consistently lead to larger
coarsening rates.

In summary, we have presented results for several independent measures of the
linear scale of stripe patterns ordering very near onset, and obtained a 
coarsening exponent that is very close to $z=3$. This value can be explained 
through dimensional 
analysis of the velocity of a single grain boundary advancing into a background
of curved stripes in the limit $\epsilon \rightarrow 0$. This mechanism also
predicts increasing corrections to scaling further from onset due to defect
pinning.

This research has been supported by the U.S. Department of Energy, contract
No. DE-FG05-95ER14566.

\bibliographystyle{prsty}
\bibliography{/a/alpha2/home/scri49/users/vinals/mss/references}

\end{document}